\documentclass[amsmat,amssymb,amsfonts,aps,prl,twocolumn,showpacs]{revtex4}

\usepackage{graphicx}
\usepackage{dcolumn}
\usepackage{bm}
\begin{document}

\title{ Colossal  anisotropy  in diluted magnetic topological insulators}
\author{ A. S. N\'u\~nez (1), J. Fern\'andez-Rossier(2)}
\affiliation{
(1)  Departamento de F\'isica, Facultad de Ciencias F\'isicas y
Matem\'aticas, Universidad de Chile, Casilla 487-3, Santiago, Chile \\
(2) Departamento de F{\`i}sica Aplicada, Universidad de Alicante, San
Vicente del Raspeig, Spain }

\date{\today} 

\begin{abstract} 
We consider  dilute magnetic doping in the surface of a  three dimensional topological insulator where a two dimensional Dirac  electron gas resides.  We find that  exchange coupling between  magnetic atoms and the Dirac  electrons has a strong and peculiar effect on both.  
First,   the  exchange-induced single ion  magnetic anisotropy is very large and favors off-plane orientation. In the case of ferromagnetically ordered phase    we find a colossal magnetic anisotropy energy, of the order of the critical temperature.
Second,    a persistent  electronic current circulates around the magnetic atom     and, in the case of a ferromagnetic phase, around the edges of the surface.

\end{abstract}

\maketitle
 
 The prediction\cite{Kane-Mele1,Kane-Mele2,Bernevig06,Zhang09} and subsequent experimental confirmation\cite{Konig07,Chen09} of   topological insulators opens an entirely new research field with fascinating possibilities\cite{Zhang2010,Review-Kane,TI}.   
 Three dimensional topological insulators, like  Bi$_2$Te$_3$ and Bi$_2$Se$_3$ , host a  helical two dimensional  electron gas  in their surfaces  governed by the  Dirac equation in which the spin of the electrons is
  strongly coupled to their momentum,   in contrast with graphene, for which the components of the Dirac spinor are associated to the atomic sublattice.  In this way,   spin-orbit coupling is the dominant term in the Hamiltonian and makes this system different from most of electron gases studied so far.

 The  helical  electron gas  has exotic and non-trivial  responses to external perturbations, like applied electric and magnetic fields,  the proximity of superconductors\cite{Zhang-monopole} and ferromagnets\cite{Nagaosa2010,Garate09}. Diluted  Magnetic doping is particularly appealing\cite{Kul01,Bi2Te3Mn,Bi2Se3Fe,ZhangRKKY,Liu2009,Biswas09,Ye2010,ZhangQAHE} and has been demonstrated experimentally\cite{Bi2Te3Mn,Bi2Se3Fe}.  On one side, it could result in bulk  ferromagnetic order coexisting with the topological order\cite{ZhangRKKY, ZhangQAHE}, very much like conventional diluted magnetic semiconductors preserve the electronic properties of the parent compounds \cite{DMS,Dietl,Abolfath,JFR01}.  On the other side, the helical electron gas  is expected to affect strongly the  spin dynamics of the  magnetic atoms.  In particular, because of the large spin-orbit coupling,  the helical electron gas might induce a large single ion magnetic anisotropy on the magnetic atoms, as it happens in the case of Co in Pt \cite{Gambardella}.  This diluted magnetic system could be gated\cite{Jarillo2010}, which could permit to control both the properties of a collective ferromagnetic phase \cite{Ohno} as well as the properties of a single magnetic atom \cite{Leger06}, and it could be probed with STM, which permits additional insight and control. 
 
  Here we study how the spin of magnetic atoms and the Dirac electron gas in the surface of a topological insulator influence each other.  For that matter, we assume that the surface of the topological insulator is weakly doped with magnetic  atoms (see figure 1). We assume that the Fermi energy is not changed. This could be achieved either with isoelectronic magnetic dopants, by controlling the charge with a gate\cite{Ohno} or by additional compensating non-magnetic dopants \cite{Hor09}.
  We find that,  due to its coupling to the Dirac electron gas,   a single magnetic ion has an additional source of  magnetic anisotropy which favors off-plane alignment. We study the  case of  a density of many magnetic dopants ferromagnetically ordered due to indirect exchange coupled mediated by the Dirac electron gas.  We  find that the magnetic anisotropy energy density of this phase  is the same than the energy gained by the system due to magnetic order. We refer to this as  colossal magnetic anisotropy, in analogy to a similar result obtained for Pt chains \cite{SMO07}. Finally, we show that magnetic dopants induce charge currents in the Dirac electron gas.

We consider the following first quantization Hamiltonian for local spins $\vec{\Omega}_I$ and Dirac electrons:
\begin{equation}
{\cal H}= v_F \vec{\tau}\cdot \vec{p}+ \frac{1}{2} \vec{\tau}\cdot
\sum_I J(|\vec{r}-\vec{r}_I|) \vec{\Omega}_I
\label{Hamil0}
\end{equation}
In the first term,  the Dirac Hamiltonian for two dimensional electrons, the electron momentum
  $\vec{p}=(p_x,p_y)$ lies in the plane. In the second term, the  spin $\vec{S}=\frac{1}{2}\vec{\tau}$  is exchanged coupled to   $\vec{\Omega}_I$,  which are classical unit vectors\cite{Liu2009} representing the spin orientation of the magnetic atom located in $\vec{r}_I$ and 
  $J(r)$  is an exchange potential of range $a$, the  interatomic distance.  We assume half-filling, i.e.,   only the low energy band is occupied.  

 \begin{figure}
[hbt]
\includegraphics[width=0.9\linewidth,angle=0]{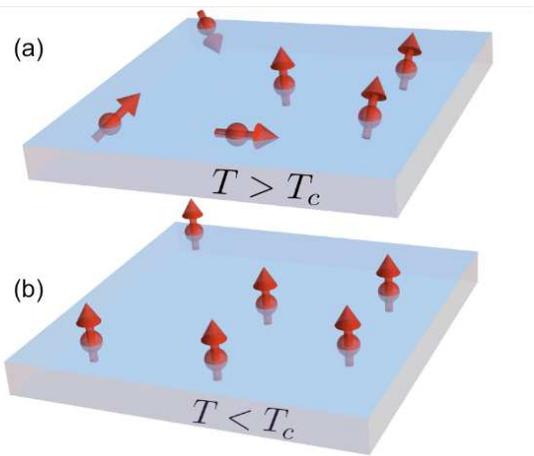}
\caption{ \label{fig1}. Color online.  Scheme of the surface of a topological insulator weakly doped with magnetic atoms, without magnetic order (top panel) and in the ferromagnetic pase (bottom panel) }
\end{figure}
\begin{figure}
[hbt]
\includegraphics[width=0.9\linewidth,angle=0]{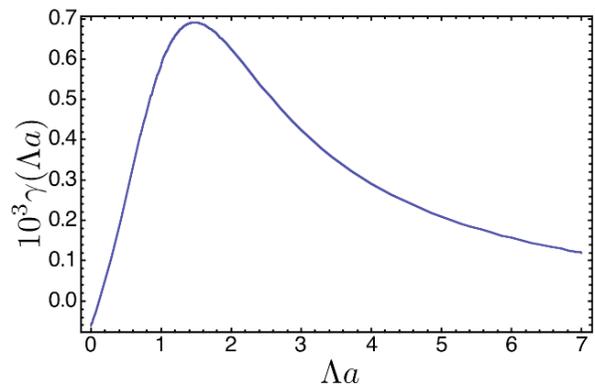}
\caption{ \label{fig2} Prefactor $\gamma(a\Lambda)$,  for the magnetic anisotropy energy for a single magnetic atom, as a function of $\Lambda a$, calculated numerically, for different values of $\Lambda$. The points collapse into a single curve  }
\end{figure}

 In the following we consider two situations: the single spin case and the many spin ferromagnetic phase in the  virtual crystal approximation \cite{DMS}.   We treat the  single spin case numerically.  For that matter, we represent the
 Hamiltonian (\ref{Hamil0}), for a single spin $\vec{\Omega}$, in the basis of eigenstates of the unperturbed Dirac Hamiltonian:
 \begin{equation}
{\bf  \Psi}_{s,\vec{k}}(\vec{r})= \frac{1}{\sqrt{2 A}} e^{i\vec{k}\cdot\vec{r}}
 \left(\begin{array}{c}
 e^{i \phi_k} \\ s\end{array}\right) \equiv \frac{1}{\sqrt{2 A}} e^{i\vec{k}\cdot\vec{r}} {\bf F}_{s\vec{k}}
 \label{spinor}
 \end{equation} 
where $A$ is the sample area,  $\vec k = |\vec{k}| (cos \phi_k, sin \phi_k)$,  $s=\pm 1$. The corresponding unperturbed energies read $\epsilon_{s}(\vec{k})= s\hbar v_F  |\vec{k}|$.   In this basis, the  Hamiltonian of a Dirac electron gas interacting with a single spin reads:
\begin{equation} 
{\cal H}_{\rm single}= \epsilon_s(\vec{k}) \delta_{s,s'}\delta_{\vec{k},\vec{k}'}  +  \frac{1}{2}J(\vec{k}-\vec{k}') 
 \vec\Omega\cdot ({\bf F}^{\dagger}_{s,\vec{k}} \vec{\tau}{\bf F}_{s',\vec{k}'})
 \label{Hamil1}
 \end{equation}
  where  $J(\vec{k})$ is the Fourier transform of $J(r)$. 
  We take $J(r)= \frac{j}{\pi a^2} e^{-(r/a)^2}$, where $j$ has dimensions of energy times area,  so that 
 $J(\vec{k}-\vec{k}')= \frac{j}{A} e^{-(|\vec{k}-\vec{k'}|a/2)^2}$.  We consider $N\simeq 900$ points in a  square  of area $\Lambda^2$ centered around $\vec{k}=(0,0)$. Here $\Lambda$ is a momentum cutoff which satisfies $\Lambda a\simeq 1$.   For a given spin orientation $\vec{\Omega}$   we obtain the $2 N$ eigenstates $E_{n}(\vec\Omega)$ of eq. (\ref{Hamil1}). At half filling the multielectron ground state energy is given by $E_G(\vec{\Omega}))=\sum^N_{n=1} E_{n}(\vec{\Omega})$ .  As expected by symmetry, we find that $E_G$ does not depend on the in-plane components $\Omega_{x,y}$, being  a function of $\Omega_z$ only. From our numerical results we infer that the single ion magnetic anisotropy is given by:
 \begin{equation}
 \delta{\cal E} = E_G(\Omega_z)- E_G(\hat{x})=  - \gamma(\Lambda a) \frac{j^2}{\hbar v_F} \Lambda ^3  \Omega_z^2
 \end{equation}
where $\gamma(\Lambda a)$ is shown in figure (\ref{fig2}). 
This is one of the main results of this manuscript: because of its coupling to the Helical electron gas,  a single magnetic atom acquires a off-plane  magnetic anisotropy.  
 We estimate the single ion magnetic anisotropy taking $j=Ja^2$, $\Lambda a\simeq1$, and, following reference \cite{ZhangRKKY},  $J=0.5 eV$ and $\hbar v_f \Lambda=0.1$ eV.  We obtain  $\delta{\cal E}\simeq$2.5 meV.  
This single ion term competes with RKKY interactions \cite{ZhangRKKY,Biswas09,Ye2010} and with the intrinsic single ion anisotropy that arises from the interplay of the $d$ electron spin orbit and their crystal field splitting.  The single ion anisotropy measured  in inelastic scanning tunneling spectroscopy \cite{Heinrich,JFR09}

  We now consider the case of many magnetic atoms. Due to the single ion anisotropy, we expect them to point perpendicular to the surface.  In that situation, the indirect exchange coupling is ferromagnetic \cite{ZhangRKKY}, so that we can expect them to order ferromagnetically below a certain critical temperature that we estimate below.  We study this ferromagnetic phase in the  virtual crystal approximation (VCA)\cite{DMS,Dietl,Abolfath,JFR01}, which is equivalent to replace the inhomogeneous and non-collinear exchange potential in  eq. (\ref{Hamil0}) by an constant exchange field,  proportional to the concentration of magnetic atoms, $c=N_{\rm imp}/A$,  whose spins all point along the same direction $\vec{\Omega}$. The effective Hamiltonian is now translational invariant and, in the momentum space, reads
\begin{equation}
{\cal H}_{\rm VCA}= \hbar v_F \vec{k}\cdot \vec{\tau} + \frac{\Delta}{2}\vec{\Omega}\cdot\vec{\tau}
\end{equation}
where $\Delta=jc$.
The off-plane  $\Omega_z$ component plays a role very different from  $\Omega_x,\Omega_y$.  The latter can be gauged away, by choosing a constant vector potential $A_b=  \frac{\Delta}{2  v_f}\Omega_b$, with $b=x,y$ (see figure 3). Since this vector potential is constant, it has no effect on the physical properties of the system. In particular,  the total energy does not depend on $\Omega_{x,y}$,  as it  happens in  the single ion case.  In contrast, the perpendicular magnetization can not be gauged away and
it has an important effect on the energy bands: it opens a gap of magnitude $\Delta$ (see figure 3).  The new energy bands read now $\epsilon_{s}(\vec{k,\Omega_z})= s \sqrt{\frac{1}{4}\Delta^2 \Omega_z+ (\hbar v_F k)^2}$. We calculate
the multielectron ground state energy, with respect to the non-magnetic ground state. To leading order in the cutoff $\Lambda$ we obtain
\begin{equation}
\frac{\delta{\cal E}}{A}  
\simeq - \frac{\Delta^2 \Omega_z^2}{2\hbar v_F} \Lambda 
\label{etotvca}
\end{equation}
This equation  can be interpreted in two different ways. On one side, this is the energy that the system gains in the ferromagnetic phase,  with perpendicular easy axis, with respect to the non magnetic phase. The minus sign denotes that the helical gas is a paramagnet, in spite of the fact that the density of states vanishes at the Fermi energy.   Equation (\ref{etotvca}) permits to derive an expression for the transition temperature. For that matter, we expand the  free energy of the magnetic atoms
 around  $\Omega_z=0$ . To  leading order, we find  
 an  entropic  term, that favors the non-magnetic phase,  and the electronic term, which favors magnetic order \cite{JFR01}:
\begin{equation}
\frac{\cal F}{A}=  c \frac{\Omega_z^2}{2\chi_0} - \frac{\Delta^2 \Lambda } {2\hbar v_F}\Omega_z^2
\end{equation}
where $\chi_0= \frac{S(S+1)}{3 k_B T}$.   At the critical temperature  the prefactor of $\Omega_z^2$ changes sign:
\begin{equation}
k_B T_c =   \frac{S(S+1)j^2 c } {6\hbar v_F} \Lambda 
\end{equation}
On the other hand, equation (\ref{etotvca}) can be interpreted as the magnetic anisotropy energy associated to a collective rotation of the magnetization from in plane to off-plane.  Thus, we can write:
\begin{equation}
\frac{\delta{\cal E}}{A}=-3c\frac{ k_B T_c}{S(S+1)}    \Omega_z^2
\label{colossal}
\end{equation}
This result shows the peculiar nature of a ferromagnetic phase in the surface of a diluted magnetic topological insulator. The magnetic anisotropy and the Curie temperature are the same.  In other words,  a rotation of the collective magnetization from its off-plane easy axis is as expensive energetically as a transition to the non-magnetic case.  Thus, a collinear magnetic phase is only possible when the magnetic moment is perpendicular to the surface. We refer to this  as colossal magnetic anisotropy, in analogy with the magnetic phase predicted for Pt nanowires \cite{SMO07}.  Equation (\ref{colossal}) points out how radically different is  the ferromagnetic phase in a diluted magnetic topological insulator compared to from the ferromagnetic order in metals and diluted magnetic semiconductors, where the  Curie temperature and magnetic anisotropy are very different \cite{Dietl,Abolfath}. 
This result arises from the fact that the band energy and the spin orbit coupling are the same in the Helical electron gas.  

 \begin{figure}
[t]
\includegraphics[width=1.0\linewidth,angle=0]{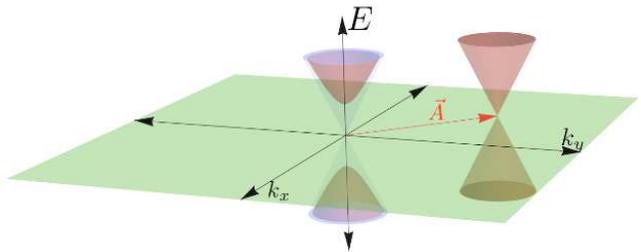}
\caption{ \label{fig3}   Color on-line. Energy bands in the non-magnetic phase  (gapless bands) and in the magnetic phase with $\vec{\Omega}=\hat{z}$ (gapped case) and $\vec{\Omega}=\hat{x}$ (shifted gapless bands) }
\end{figure}

We now estimate the Curie temperature for a diluted magnetic topological insulator. For that matter we consider Bi$_2$Te$_3$, doped with iron, for which $S=2$.  We assume $j=J a^2$  so that we can write 
$k_b T_c =   \frac{S(S+1)J^2 ca^2 } {6\hbar v_F} \Lambda a^2$.  We assume $\Lambda a =1$ and $\hbar v_F/a=0.1$ eV.
For $J=0.5 eV$ and  $ca^2=0.01$,  we obtain $k_b T = 25 $meV, i.e., room temperature.  Notice that this prediction scales with $J^2$, so that a 10-fold reduction of $J$  results in a 100-fold reduction of $T_c$.  In addition, a mean field prediction of $T_c$ is likely to overestimate the real transition temperature.  

We now consider the effect of the magnetic moments on  the Helical electron gas.  It has been shown that a single magnetic impurity in the surface of a topological insulator opens a gap in the local density of states \cite{ZhangRKKY,Biswas09} and, in the case of the  magnetic spin pointing perpendicular to the surface,  it  generates a circulating in-plane spin density around the magnetic atom.   Importantly, the spin and the current operators are the same, modulo a constant, in the case of Dirac electrons:
\begin{equation}
\vec{j}\equiv \frac{\delta }{\delta \vec{A}}{\cal H}(\vec{p}-e\vec{A})= -ev_F \vec{\sigma}
\label{spinequalcurrent}
\end{equation}
In  consequence,  the in-plane circulating spin density obtained in references (\onlinecite{ZhangRKKY,Biswas09}) can be also interpreted as a circulating charge current around the magnetic impurity.  In turn, the circulating charge current bears an orbital magnetic moment perpendicular to the plane.  Thus the off-plane spin  induces an off-plane orbital moment.  If two of such magnetic atoms are close to each other and co-polarized, the sum of their charge currents is such that there is no current flow in between them, but there is current flowing around them.  The argument can be extended to the entire surface. Thus, we expect that in the edges of the surface of a ferromagnetically ordered diluted magnetic topological insulator there is a circulating charge current in the ground state.  This persistent current induced by a space dependent magnetic order parameter is similar to the one recently predicted for  graphene with magnetic order in the edges \cite{Soriano2010}. 

 We now provide a general argument to show that, in the presence of a off-plane  varying spin density, the Dirac electrons must have an in-plane charge current.  We consider the  Dirac electrons in the presence of an exchange potential ${\cal V}= \frac{J}{2} \Omega_z(x)\tau_z$.  Thus, the components of the eigenstates ${\bf \Psi}(x,y)$ can be written as  $e^{i k_y y} \phi_{\sigma} (x)$.  It can be easily seen that, for a state ${\bf \Psi}_{k_y}(x,y)$ with energy $E$ and momentum $k_y$ the following identity is satisfied, for an arbitrary $\Omega_z(x)$:
\begin{equation}
2E {\cal S}_y(x) + \hbar v_F k_y n(x)  = \hbar v_F \partial_x {\cal S}_z(x)
\end{equation}
where ${\cal S}_a(x)\equiv\frac{1}{2} \sum_{\sigma,\sigma'}\phi_{\sigma}^*(x)\tau^a_{\sigma,\sigma'}\phi_{\sigma'}(x)$ and
$n(x)\equiv \sum_{\sigma} \phi_{\sigma}^*(x)\phi_{\sigma}(x)$ are the 
spin and charge densities associated to ${\bf \Psi}_{k_y}(x,y)$, respectively. Since a given state with energy $E$ and momentum $k_y$ has a degenerate partner with momentum $-k_y$,  when summing over all the states with the same energy $E$, and using equation (\ref{spinequalcurrent}), we are left with the equation:
\begin{equation}
\langle {\cal J}_y(x)\rangle_E = \frac{\hbar }{2 e E} \langle \partial_x {\cal S}_z(x)\rangle_E
\end{equation}
where $\langle\rangle_E$ stands for sum over all the states with energy $E$.  Thus,  as we move from outside the sample $x<0$ to the magnetic sample $x>0$, there is a finite  $\langle \partial_x {\cal S}_z(x)\rangle_E$ which implies current flow perpendicular to the sample boundary.  A similar analysis accounts for the current vortices induced by a single impurity.

In conclusion, we study a three dimensional topological insulator with diluted magnetic atoms in its surface. We  predict that magnetic dopants exchanged coupled to the Dirac electron gas in the surface of a topological insulator acquire a magnetic anisotropy that favors off-plane orientation. We find that the magnetic anisotropy energy and the energy gain due to magnetic order are the same.  As a result of this colossal magnetic anisotropy, the ferromagnetic phase can only exist when the spins are oriented perpendicular to the plane.   We find that magnetic dopants induce in-plane charge currents. In the ferromagnetic phase,  charge current flows in the edges of the sample.

This work has been financially supported by MEC-Spain (Grant Nos.
MAT07-67845 and CONSOLIDER CSD2007-0010), by Proyecto de Iniciaci\'on en Investigaci\'on
Fondecyt 11070008  and by N\'ucleo Cient\'ifico Milenio ``Magnetismo B\'asico
y/o Aplicado'' P06022-F.  ASN acknowledges  funding from  Universidad de Alicante.

\end{document}